\documentclass[preprint,superscriptaddress,amsmath,amssymb]{revtex4}

\usepackage{graphicx}
\usepackage{dcolumn}
\usepackage{bm}
\usepackage{ulem}
\usepackage{epstopdf}
\usepackage{amsmath}
\usepackage{amssymb}

\newcommand{\nc}{\newcommand}
\nc{\be}{\begin{eqnarray}}
\nc{\ee}{\end{eqnarray}}
\nc{\bea}{\begin{eqnarray}}
\nc{\eea}{\end{eqnarray}}
\nc{\bean}{\begin{eqnarray*}}
\nc{\eean}{\end{eqnarray*}}

\preprint{}

\begin{document}

\title{Reversed anisotropy of the in-plane resistivity in the antiferromagnetic phase of iron tellurides}

\author{L.~Liu}
\affiliation{Department of Physics, University of Tokyo, Tokyo 113-0033, Japan}

\author{T.~Mikami}
\affiliation{Department of Physics, University of Tokyo, Tokyo 113-0033, Japan}

\author{M.~Takahashi}
\affiliation{Department of Physics, University of Tokyo, Tokyo 113-0033, Japan}

\author{S.~Ishida}
\affiliation{National Institute of Advanced Industrial Science and Technology, Tsukuba 305-8568, Japan}

\author{T.~Kakeshita}
\affiliation{Department of Physics, University of Tokyo, Tokyo 113-0033, Japan}

\author{K.~Okazaki}
\affiliation{Department of Physics, University of Tokyo, Tokyo 113-0033, Japan}

\author{A.~Fujimori}
\affiliation{Department of Physics, University of Tokyo, Tokyo 113-0033, Japan}

\author{S.~Uchida}
\affiliation{Department of Physics, University of Tokyo, Tokyo 113-0033, Japan}




\begin{abstract}

We systematically investigated the anisotropic in-plane resistivity of the iron telluride including three kinds of impurity atoms: excess Fe, Se substituted for Te, and Cu substituted  for Fe. Sizable resistivity anisotropy was found in the magneto-structurally ordered phase whereas the sign  is opposite ($\rho_a$ $>$ $\rho_b$, where the $b$-axis parameter is shorter than the $a$-axis one) to that observed in the transition-metal doped iron arsenides ($\rho_a$ $<$ $\rho_b$). On the other hand, our results demonstrate that the magnitude of the resistivity anisotropy in the iron tellurides is correlated with the amount of impurities, implying that the resistivity anisotropy originates from an exotic impurity effect like that in the iron arsenides. This suggests that the anisotropic carrier scattering by impurities is a universal phenomenon in the magneto-structurally ordered phase of the iron-based materials.

\end{abstract}

\maketitle


\section{introduction}

Almost all of the parent compounds of iron-based superconductors (FeSCs) harbor a metallic antiferromagnetic (AFM) ground state \cite{Fisher_RPP11,Paglione_NP10}. In order to understand the mechanism of the appearance of superconductivity in the phase diagram of FeSCs, it is pivotal to first clarify the nature of the metallic AFM state. The AFM order in the parent compounds of iron pnictides, e.g., BaFe$_2$As$_2$, is a collinear type with spins aligning antiferromagnetically in one direction ($a$ axis) and ferromagnetically in the other ($b$ axis). Therefore, the AFM order should break the fourfold (tetragonal) symmetry of the underlying lattice of the high-temperature phase. It has been revealed that the appearance of antiferromagnetism with decreasing temperature is either accompanied or preceded by a tetragonal-to-orthorhombic structural phase transition. Intensive studies of the iron arsenides performed by transport measurements~\cite{Chu_science10,Blomberg_PRB11,Ishida_PRL13,Ishida_JACS13,Kuo_PRB11,Ying_PRL11}, optical measurements~\cite{Nakajima_PNAS11,Nakajima_PRL12}, and angle-resolved photoemission spectroscopy (ARPES)~\cite{Yi_PNAS11,Kim_PRB11}, have revealed the unique feature of the AFM ground state with unprecedented in-plane electronic anisotropy. Large resistivity anisotropy has been observed in the early studies of the AFM ground state of Co-doped $A$Fe$_2$As$_2$ (abbreviated as 122) compounds, where $A$ denotes an alkali-earth-metal element~\cite{Chu_science10}. The origin of the resistivity anisotropy has been discussed in terms of the anisotropy of the reconstructed Fermi surfaces (FSs)~\cite{Fisher_RPP11,Kuo_PRB11} or orbital ordering~\cite{Chen_PRB10} in the orthorhombic phase. However, from recent studies of annealed Co-doped 122 compounds with significantly improved qualities, an impurity-induced-anisotropy scenario was proposed to explain the observed resistivity anisotropy~\cite{Ishida_PRL13,Nakajima_PRL12}. Those studies provided evidence that a doped Co atom forms an impurity state which scatters carriers anisotropically.

\begin{figure}[htb]
\begin{center}
\includegraphics[width=8.5cm]{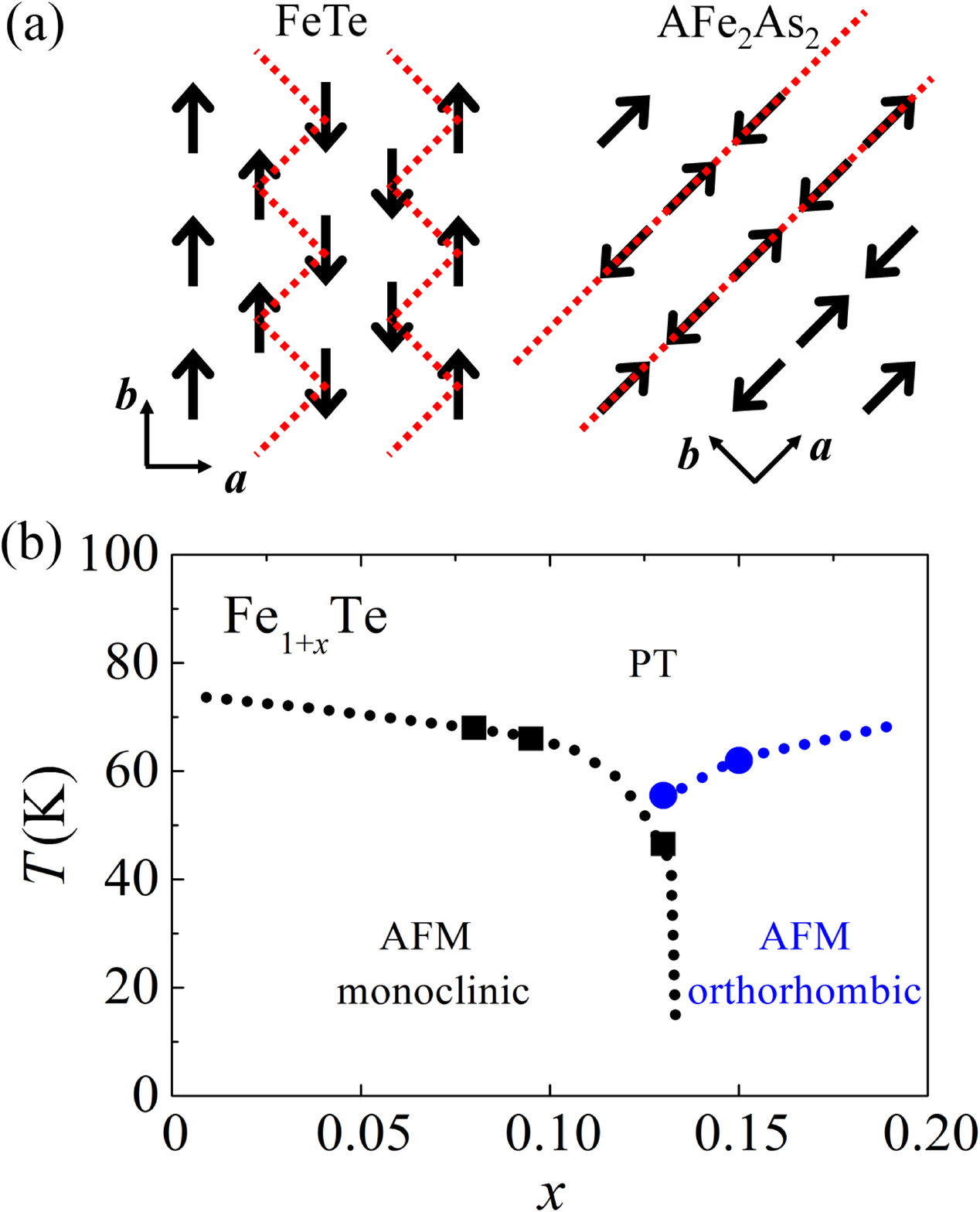}
\caption{\label{fig1} (color online) (a) AFM order of FeTe and AFe$_2$As$_2$ within the $ab$ plane. The definition of $a$- and $b$-axes in the AFM state is indicated. (b) Phase diagram of Fe$_{1+x}$Te. The magneto-structural transition temperature ($T_s$) is determined from the temperature dependence of resistivity as indicated by arrows in Fig. \ref{fig2}(a). PT denotes the paramagnetic-tetragonal phase.}
\end{center}
\end{figure}

\indent The ``parent" compound of the iron chalcogenide superconductors Fe$_{1+x}$Te (abbreviated as 11) also has an AFM ground state with broken fourfold rotational symmetry, but shows bicollinear magnetic order with a different direction of the magnetic wave vector (rotated by $45^{\circ}$ in the $ab$ plane) from that of BaFe$_2$As$_2$~\cite{Li_PRB09}. Here, it is necessary to indicate again the definition of $a$- and $b$-axes in the AFM state. For both 122 and 11 materials, the $a$-axis is defined as the longer axis with AFM spin alignment while the $b$-axis is the shorter axis with ferromagnetic spin alignment, as shown in Fig. \ref{fig1}(a). Since the directions of the AFM ordering wavevectors in the 122 and 11 materials are  different by $45^{\circ}$, the defined $a$- and $b$-axes are $45^{\circ}$ rotated from each other. Due to the similarities and differences in the two systems, extending the resistivity anisotropy measurement to the 11 system would shed light on the nature of the AFM phase and the origin of the resistivity anisotropy observed in the iron arsenides.

\indent In Fe$_{1+x}$Te, the interstitial sites of Te layers usually allow the partial occupation of iron atoms~\cite{Liu_PRB09}, resulting in the existence of excess Fe in the chemical formula, which are thought to have a large effect on the physical properties. The effect of excess Fe on transport properties has to be clarified.

\indent In this paper, we demonstrate the appearance of in-plane resistivity anisotropy in the AFM phase of iron tellurides by detwinning crystals with uniaxial pressure. Furthermore, we demonstrate correlation between the magnitude of the resistivity anisotropy and the amount of impurities (excess Fe, substituted Se and Cu atoms) in the 11 materials. The present result evidences that the impurity-induced-anisotropy scenario recently proposed for the iron arsenides also explains the resistivity anisotropy in the 11 system. Unexpectedly, the resistivity anisotropy in 11, $\rho_a$ $>$ $\rho_b$, is opposite to that observed for transition-metal doped 122, $\rho_b>\rho_a$, irrespective of the species of impurity atoms/sites.


\section{experimental}

Single crystals of Se- and Cu-substituted  Fe$_{1+x}$Te single crystals were grown by the Bridgman method~\cite{Wen_RPP11}. The actual composition of the samples were determined by energy-dispersive X-ray spectroscopy (EDXS) analysis. A standard four-terminal method was used for the in-plane resistivity measurements on twinned crystals. In order to measure the intrinsic in-plane resistivity anisotropy hampered by the twin formation below the magneto-structural transition temperature ($T_s$), samples need to be detwinned. Samples were cut into a rectangular shape with the larger face in the cleaved $ab$ plane, the edges of which are along the tetragonal $\left\langle100\right\rangle$ axes and would become $a$ or $b$ axes in the AFM phase. Samples can be effectively detwinned by applying compressive pressure along the edge, in which the shorter $b$ axis would be favored. We used the Montgomery method to measure the resistivity along the $a$ and $b$ axes simultaneously. More details of the uniaxial pressure cell have been described elsewhere~\cite{Liang_JPCS11}. The measurements were performed in a Quantum Design Physical Property Measurement System (PPMS) and all the data were obtained while warming the samples.


\section{Results and Discussions}

\subsection{Effects of excess Fe in  Fe$_{1+x}$Te}

It has been clarified through the neutron diffraction~\cite{Rodriguez_PRB11} and X-ray diffraction measurements~\cite{Mizuguchi_SSC12} that the excess Fe atoms affect the crystal structure and magnetic ordering in the low-temperature phase of  Fe$_{1+x}$Te. The ordering vector and the crystal structure exhibit a crossover around a critical content $x_c$ of excess Fe. The low-temperature phase in the region of $x < x_c$ is monoclinic with a commensurate AFM ordering, while it is orthorhombic with an incommensurate AFM ordering in the region of high excess Fe concentration. From our resistivity measurement, $T_s$ is determined by the temperature at which the derivative of the resistivity curve shows a peak feature, like the method used in Ref. \cite{Chu_PRB09}. The phase diagram of  Fe$_{1+x}$Te thus obtained and illustrated in Fig. \ref{fig1}(b) is consistent with the previous result obtained from the magnetic susceptibility measurements~\cite{Mizuguchi_SSC12}.

\begin{figure}[htb]
\begin{center}
\includegraphics[width=8.5cm]{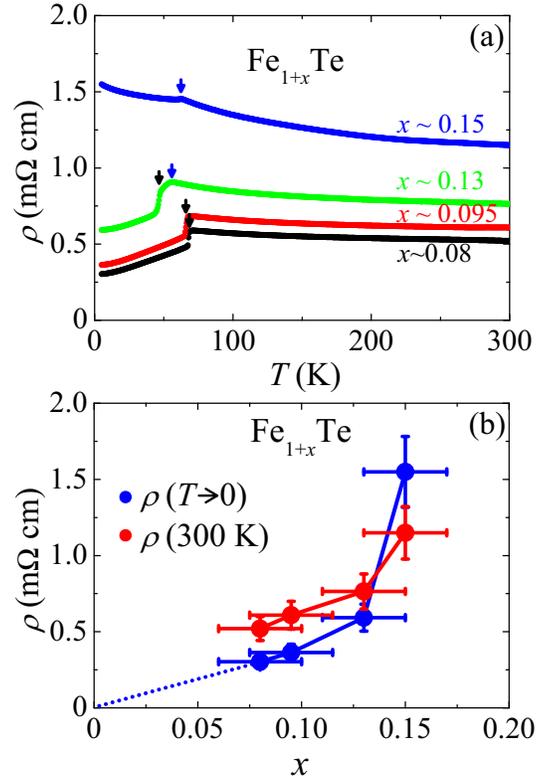}
\caption{\label{fig2} (color online) Temperature dependence of the in-plane resistivity of Fe$_{1+x}$Te with different excess Fe content $ x $. (b) Magnitude of the residual resistivity and the resistivity at 300 K plotted against the excess Fe content $x$. }
\end{center}
\end{figure}

\indent   Figure \ref{fig2}(a) shows the temperature ($T$) dependence of the in-plane resistivity for the samples with different excess Fe contents in the range between $x$ $\sim$ 0.08 and 0.15. For $x$ $\sim$ 0.08, the $T$ dependence is characterized by a discontinuous jump at 68 K, corresponding to the tetragonal-to-monoclinic structural transition accompanied by the paramagnetic-to-AFM transition. The $T$ dependence of resistivity is divided into two distinct regions by the phase transition: in the paramagnetic-tetragonal (PT) phase, it is semiconductor-like whereas in the AFM-monoclinic phase, resistivity shows a metallic behavior. In the perspective of spectroscopy measurements~\cite{Zhang_PRB10}, the ARPES spectra near $E_F$ are characterized by a very broad feature in the PT phase, indicating a highly incoherent character of carriers and consequently causing a semiconductor-like transport behavior in the resistivity. This is also consistently reflected in the optical conductivity spectra~\cite{Chen_PRB09}. No well-defined Drude component is observed above $T_s$ while a Drude component develops below $T_s$. For $x \leq 0.13$, the $T$ dependence of the resistivity is mostly similar while $T_s$ slightly decreases with increasing excess Fe content. For $x$ $\sim$ 0.13, the $T$ dependence of the in-plane resistivity shows a two-step feature in the transition. One can see in the phase diagram that this composition corresponds to the region around $x_c$ where two phase transitons occur sequentially. X-ray diffraction measurements have shown that in this region the tetragonal lattice first distorts into orthorhombic and then mostly becomes monoclinic with further cooling~\cite{Mizuguchi_SSC12}, which is reflected in the $T$ dependence of resistivity with the two-step transition. For $x$ $\sim$ 0.15, the $T$ dependence of resistivity evolves into a kink feature at $T_s$ $\sim$ 62 K. In this case, both the PT phase and AFM phase show semiconducting behaviors. It is likely that the coherent Drude component observed in the low-temperature phase of  Fe$_{1+x}$Te with low $x$ disappears and all the carriers become incoherent even in the low-temperature phase due to strong disorder.

\indent There is another important feature in the magnitude of the resistivity. One can clearly see in Fig. \ref{fig2}(b) that the magnitude of the resistivity monotonically increases with excess Fe content both in the PT phase and AFM phase, indicating the strong scattering character of excess Fe. In the well-studied 122 system, the impurities, e.g., dopants added into the parent compound, usually show strong elastic scattering in the AFM phase but weak scattering in the PT phase \cite{Ishida_PRL13}. In contrast, excess Fe exhibits strong scattering in both phases. The strong scattering effect in the PT phase is probably responsible for the detrimental effect of excess Fe on the superconductivity in  Fe$_{1+x}$Te$_{1-y}$Se$_y$ as suggested in Ref.~\cite{Bendele_PRB10}. For $x \leq 0.13$, the residual resistivity (RR) shows a good linear relationship with $x$, which can be extrapolated roughly to the origin. This implies that the excess Fe acts as an elastic impurity scattering center for the carriers in the AFM phase. When  $x$ exceeds the critical value $x_c$ $\sim$ 0.13, a rapid increase in RR with Fe content is observed. It appears that excess Fe changes its character when $x$ exceeds the critical value, where the effect of disorder becomes further enhanced.

\subsection{Effects of Se and Cu substitution in  Fe$_{1+x}$Te}

\begin{figure}[htb]
\begin{center}
\includegraphics[width=8.5cm]{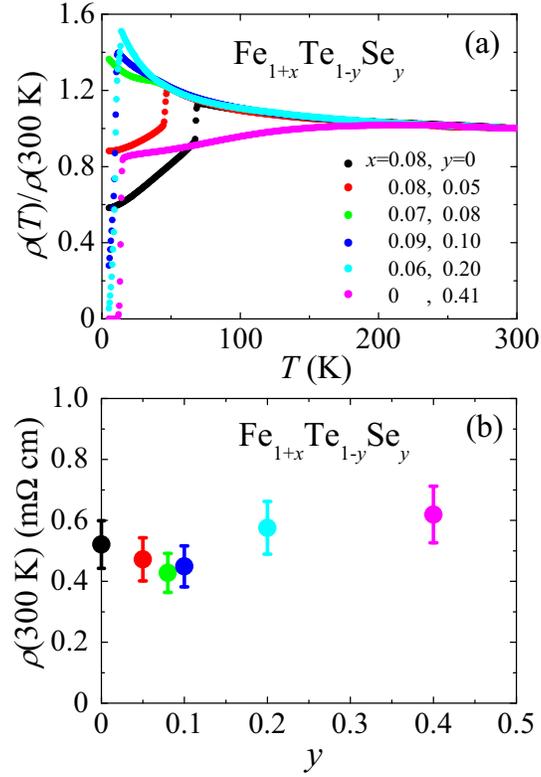}
\caption{\label{fig3} (color online) In-plane resistivity for Fe$_{1+x}$Te$_{1-y}$Se$_y$. (a) Evolution of the temperature dependence of the normalized in-plane resistivity. (b) Magnitude of the resistivity at 300 K plotted against Se content $ y $.}
\end{center}
\end{figure}

Figure \ref{fig3}(a) shows the evolution of the $T$ dependence of the normalized in-plane resistivity for Fe$_{1+x}$Te$_{1-y}$Se$_y$. The transition temperature $T_s$ systematically decreases upon Se substitution. In the AFM phase, the metallic $T$ dependence of resistivity eventually evolves into a semiconducting behavior. For $y$ = 0.08, instead of the discontinuous jump, the resistivity shows a kink feature at $T_s$. In the PT phase, there is no discernible change in the $T$ dependence for $y \leq  0.2$ while a weakly metallic behavior appears for heavily doped $y$ = 0.41. This can be naturally expected since the end material FeSe shows a good metallic behavior~\cite{Mizuguchi_JPSJ10}. Figure \ref{fig3}(b) illustrates $\rho$(300K) of Fe$_{1+x}$Te$_{1-y}$Se$_y$. No systematic change of $\rho$(300K) can be seen upon Se substitution unlike the case of isovalent P-doped 122 materials where carriers become more coherent with P substitution and resistivity consequently decreases due probably to a chemical pressure effect~\cite{Nakajima_SR14}.

\begin{figure}[htb]
\begin{center}
\includegraphics[width=8.5cm]{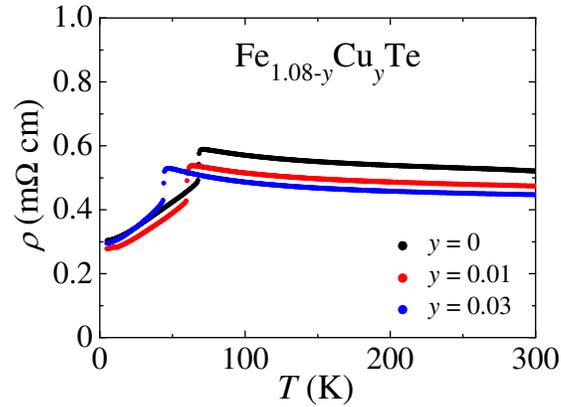}
\caption{\label{fig4} (color online) Evolution of the temperature dependence of the in-plane resistivity with Cu substitution for Fe$_{1.08-y}$Cu$_y$Te ($y \leq 0.03$).}
\end{center}
\end{figure}

\indent Figure \ref{fig4} shows the evolution of the $T$ dependence of in-plane resistivity of Cu-substituted Fe$_{1+x}$Te. Analysis of composition by EDAX indicated that the content of excess Fe was always around  $x$ $\sim$ 0.08 and that $y = 0$, 0.01, and 0.03. Upon Cu substitution, $T_s$ is systematically reduced from $T_s$ = 68 K for Cu-free samples to $T_s$ = 44 K for  $y = 0.03$. This observation suggests that Cu atoms effectively substitute for in-plane Fe atoms, causing the reduction of the ordering temperature probably due to magnetic dilution by Cu ions with smaller magnetic moment. Contrary to the strong scattering effect of excess Fe on the absolute value of resistivity, the resistivity shows small change in the metallic region upon Cu substitution, while it becomes slightly smaller in the PT phase. It has been revealed that doped transition-metal atoms in 122 systems act to strongly increase the elastic scattering rate in the AFM phase~\cite{Ishida_JACS13}. In the iron telluride, the contribution to RR purely from elastic scattering by Cu atoms could be masked by the incoherent component in the AFM state which also has a large contribution to conductivity.

\begin{figure}[htb]
\begin{center}
\includegraphics[width=8.5cm]{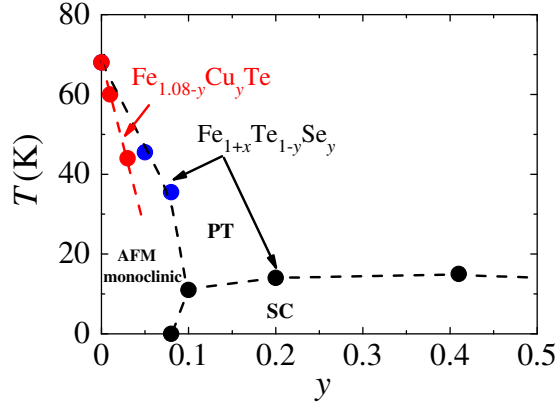}
\caption{\label{fig5} (color online) Phase diagrams of Fe$_{1+x}$Te$_{1-y}$Se$_y$ and Fe$_{1.08-y}$Cu$_y$Te ($y \leq 0.03$) based on the in-plane resistivity measurements.}
\end{center}
\end{figure}

\indent Based on our resistivity measurements, the phase diagrams of Fe$_{1+x}$Te$_{1-y}$Se$_y$ and Fe$_{1.08-y}$Cu$_y$Te are illustrated in Fig. \ref{fig5}. The obtained phase diagram of Fe$_{1+x}$Te$_{1-y}$Se$_y$ here is consistent with previous studies~\cite{Dong_PRB11,Noji_JPSJ10}. One can also see that Cu substitution suppresses $T_s$ more strongly than the isovalent Se substitution, as in the case of 122 materials.

\subsection{In-plane resistivity anisotropy in the magneto-structurally ordered phase}

\begin{figure}[htb]
\begin{center}
\includegraphics[width=8.5cm]{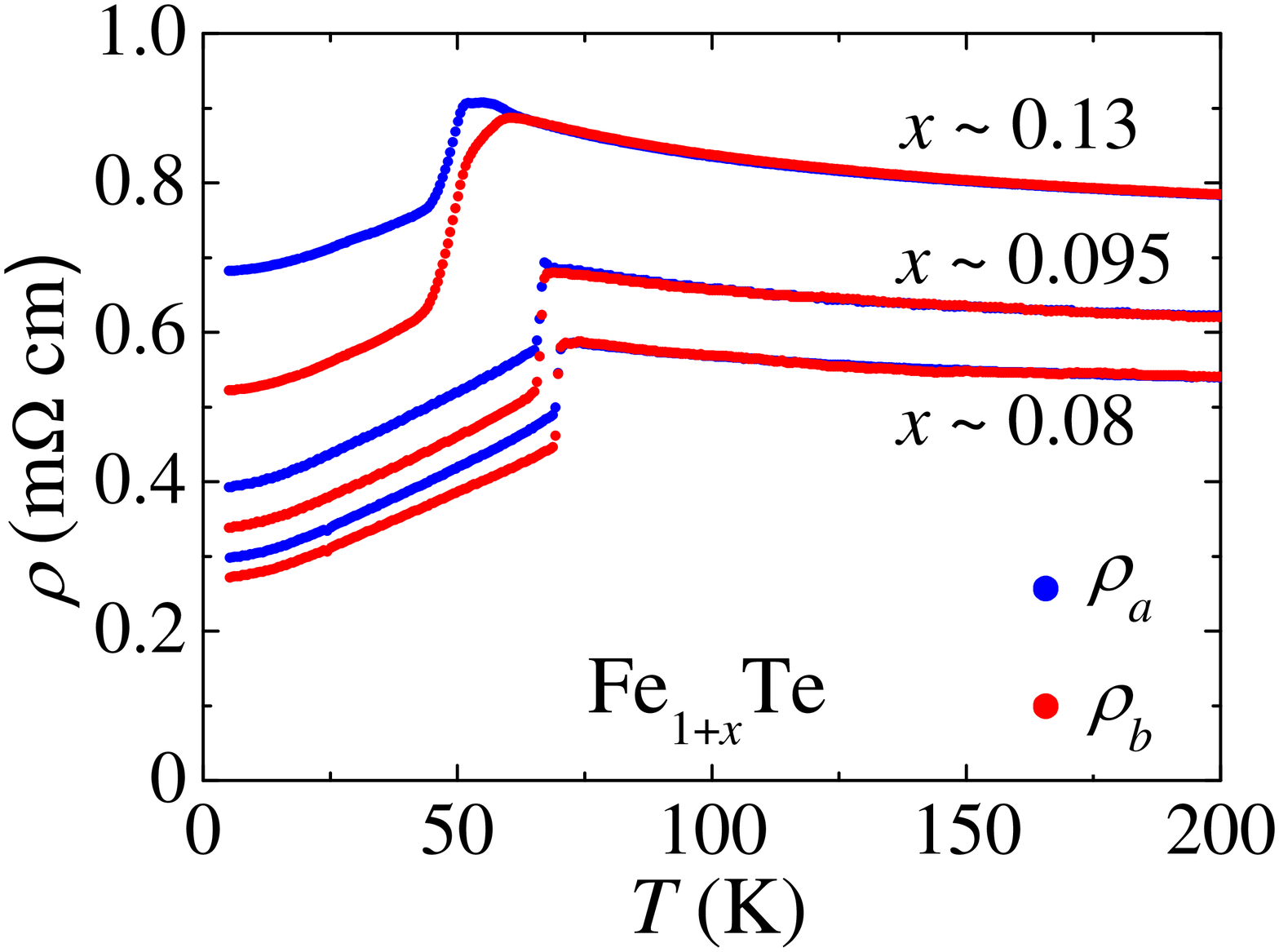}
\caption{\label{fig6} (color online) Temperature dependence of the in-plane resistivity anisotropy measured on detwinned Fe$_{1+x}$Te crystals with three different excess Fe contents.}
\end{center}
\end{figure}

\begin{figure}[htb]
\begin{center}
\includegraphics[width=8.5cm]{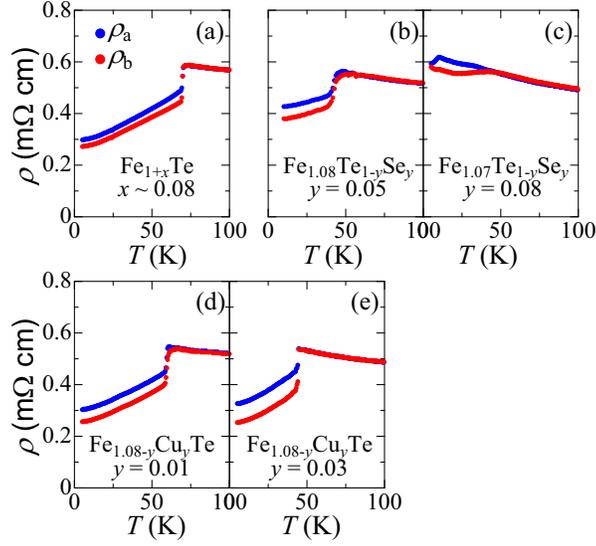}
\caption{\label{fig7} (color online) Temperature dependence of the in-plane resistivity anisotropy measured on detwinned crystals: (a) Fe$_{1.08}$Te; (b) Fe$_{1.08}$Te$_{1-y}$Se$_y$ ($y = 0.05$); (c) Fe$_{1.07}$Te$_{1-y}$Se$_y$ ($y = 0.08$); (d) Fe$_{1.08-y}$Cu$_y$Te ($y = 0.01$); (e) Fe$_{1.08-y}$Cu$_y$Te ($y = 0.03$).}
\end{center}
\end{figure}

The in-plane resistivity anisotropies of  Fe$_{1+x}$Te single crystals with three different excess Fe contents, which show the commensurate AFM ordering at low temperatures, were investigated. The results are shown in Fig. \ref{fig6}. No anisotropy was observed well above $T_s$, as in the case of SrFe$_2$As$_2$ and CaFe$_2$As$_2$, whose phase transition is also essentially of the first order. Anisotropy suddenly sets in at $T_s$. In every case, the magnitude of the anisotropy $\lvert\rho_a - \rho_b\rvert$  does not change much with lowering temperature below $T_s$. Notably, the resistivity along the $a$ axis ($\rho_a$) is higher than that along the $b$ axis ($\rho_b$), $\rho_a$ $>$ $\rho_b$, that is, the resistivity in the direction of the longer axis and antiferromagnetic spin alignment is larger than that in the direction of the shorter axis and ferromagnetic spin alignment. The result is consistent with the results by Jiang \textit{et al.} on  Fe$_{1+x}$Te with $x$ $\sim$ 0.088~\cite{Jiang_PRB13}. Note that this anisotropy is opposite to that observed in the transitional-metal doped iron arsenides. By analogy with the study of iron arsenides, we also studied the resistivity anisotropy of the isovalent Se- and heterovalent Cu-substituted  Fe$_{1+x}$Te. The results are shown in Fig. \ref{fig7}. In both cases, the resistivity anisotropy appearing in the AFM phase with $\rho_a$ $>$ $\rho_b$ always retains the same as that in the parent compounds  Fe$_{1+x}$Te, while opposite to that in the iron arsenides.

\indent It should be noted that $\rho_a$ and $\rho_b$ show similar temperature dependences in the AFM phase for all the measured samples except for that with 8$\%$ Se substituted. This means that the resistivity anisotropy is determined mostly by the anisotropy of the temperature-independent RR component. For Fe$_{1+x}$Te$_{1-y}$Se$_y$ with $y = 0.08$ [Fig. \ref{fig7}(c)], the AFM phase is strongly suppressed by disorder introduced by Se, evidenced by the insulating behavior below $T_s$. Consequently, the character of the first-order transition is weakened, resulting in the gradual increase of resistivity anisotropy below $T_s$. Below 10 K, the resistivity anisotropy suddenly becomes small due to the appearance of possible filamentary superconductivity. Moreover, for Se substitution, the transition is somewhat rounded and, therefore, the resistivity anisotropy no longer shows a sharp onset.

\begin{figure}[htb]
\begin{center}
\includegraphics[width=8.5cm]{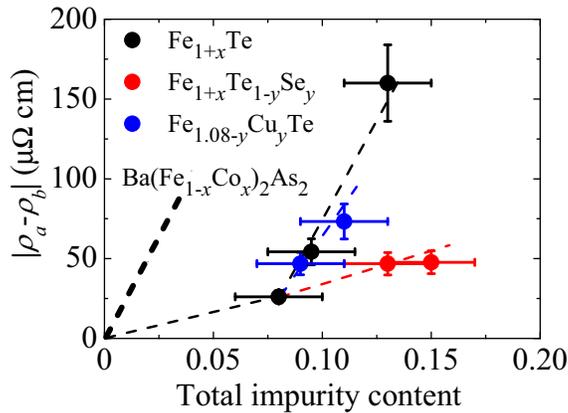}
\caption{\label{fig8} (color online) In-plane resistivity anisotropy in the residual component ($\lvert\rho_a - \rho_b\rvert$) plotted against the total impurity content: $ x $ for Fe$_{1+x}$Te, ($ x +y$) for Fe$_{1+x}$Te$_{1-y}$Se$_y$, and ($0.08+y$) for Fe$_{1.08-y}$Cu$_y$Te. The thick dashed line illustrates the dependence of the resistivity anisotropy on the substituted Co content $ x $ for Ba(Fe$_{1-x}$Co$_x$)$_2$As$_2$, reproduced from Ref.~\cite{Ishida_PRL13}.} 
\end{center}
\end{figure}

\indent Furthermore, we consider the composition dependence of the magnitude of the resistivity anisotropy. In Fig.~\ref{fig8} is plotted the magnitude of the resistivity anisotropy in the RR component ($\Delta\rho = \lvert\rho_a - \rho_b\rvert$) against the total impurity content for  Fe$_{1+x}$Te, Fe$_{1+x}$Te$_{1-y}$Se$_y$, and Fe$_{1.08-y}$Cu$_y$Te, respectively. For the parent compounds  Fe$_{1+x}$Te, $\Delta\rho$ increases with the amount of excess Fe, while excess Fe atoms induce a weak decrease in the AFM ordering temperature, implying that the absolute value of resistivity anisotropy has close relationship with the existing excess Fe atoms. For samples with Cu substitution for Fe, the magnitude of the resistivity anisotropy appears to increase monotonically with substitution. In the case of Se substitution, it appears that an increase in $\Delta\rho$ is much weaker than the other two cases.

\indent For comparison, we recall the results for the Co-doped 122 materials. Systematic studies using annealed samples demonstrate that resistivity anisotropy increases linearly with the Co composition in the “pure” AFM phase~\cite{Ishida_PRL13}. Analysis of the Drude component in the optical conductivity suggests that the anisotropy in the resistivity originates from the anisotropic carrier scattering rate rather than the anisotropic effective mass~\cite{Nakajima_PRL12}. These results support the scenario that the doped Co atoms act as anisotropic scattering centers. Scanning-tunneling-spectroscopy (STS) measurements of Ca(Fe$_{1-x}$Co$_x$)$_2$As$_2$ provided a direct observation of the formation of anisotropic dopant-induced impurity states, showing the shape of $a$-axis aligned electronic dimmers~\cite{Allan_NP13}. It was found that these impurity states scatter quasi-particles in a highly anisotropic manner with the maximum scattering cross-section concentrated along the $b$ axis of Ca(Fe$_{1-x}$Co$_x$)$_2$As$_2$.

\indent The ordering temperature is suppressed with any type of disorder in the iron telluride, so that the intrinsic anisotropy of the electronic states in the ordered phase, arising from bicollinear AFM and tetragonal-symmetry breaking lattice distortion, is expected to be weakened. On the other hand, the magnitude of the resistivity anisotropy with reversed sign increases in all the three cases. Thus, the anisotropic resistivity seems unlikely to arise directly from the intrinsic electronic anisotropy, e.g., anisotropic effective mass, or the anisotropic spin structure. In other words, these results provide evidences for the extrinsic origin of the resistivity anisotropy in 11 materials. That is, the resistivity anisotropy is closely correlated with the impurity effect, similar to the case of 122 materials. For  Fe$_{1+x}$Te, as discussed above, it is excess Fe that might scatter carriers anisotropically and thereby induces the resistivity anisotropy. For Cu- and Se-substituted  Fe$_{1+x}$Te, it cannot be completely ruled out that the excess Fe content increases with Se/Cu doping, because the Fe content by the EDAX analysis is subject to an uncertainty of $\pm$ 0.02, although the analysis indicates no significant change of $x$. Given a large increase in the magnitude of the resistivity anisotropy $\Delta\rho$ with a small increase of excess Fe content, as shown in Fig.~\ref{fig8}, it might be possible that the observed change is due to a tiny change in $x$ within the EDAX error bars. However, in view of the systematic decrease with Cu-substitution and insensitiveness to the Se-substitution in the magnitude of the PT phase resistivity which increases with the excess Fe content, we consider that the change of the excess Fe content is not large enough to significantly affect the resistivity. Therefore, the observed increase of the resistivity anisotropy in the Cu- and Se-substituted 11 crystals is likely to arise largely from the chemical substitutions. The present results suggest that the Cu and Se atoms also act as anisotropic scattering centers in the ordered phase. Note that the sign of the resistivity anisotropy in the iron tellurides is always opposite to that in the iron arsenides irrespective of dopant site. This implies that around each excess Fe atom and substituted Cu/Se atom, exotic anisotropic impurity states are formed through an anisotropic  polarization of its electronic environment, and the polarization cloud is oriented along the crystallographic $b$ axis (in the case of the iron arsenide its orientation is along the $a$ axis). From the present results and discussions, the impurity scenario may be the common phenomenological origin to explain the resistivity anisotropy in both 11 and 122 compounds. Moreover, although there might be some contributions from a slight increase of the excess Fe content to the observed increase in $\Delta\rho$, a stronger carrier scattering from Cu impurity atoms than that from substituted Se atoms is evident in the plot of Fig.~\ref{fig8}. The isovalent P substitution in the 122 materials has also a much weaker effect on $\Delta\rho$ than the heterovalent (transition-metal, e.g., Co) substitution~\cite{Ishida_JACS13}.

\indent   The most striking aspect of the results described above is the sign of the resistivity anisotropy. In 122 compounds, $\rho_b$ is larger than $\rho_a$ while it is opposite in 11 systems. Our results suggest that the observed magnitude of the resistivity anisotropy in the 11 materials is also related to the formation of an anisotropic impurity state. Under the impurity scenario, the unique AFM ground state with intrinsic electronic anisotropy provides the stage where anisotropic impurity states are formed and thereby are responsible for the anisotropic scattering rate. Thus, the reason for the opposite anisotropy in the two families of materials possibly lies in the different ground state. It was theoretically proposed that, for Co-doped BaFe$_2$As$_2$ an impurity-induced local orbital order with broken C4 symmetry, which can result in the sizable resistivity anisotropy, develops in the presence of strong orbital fluctuations near $T_s$ \cite{Inoue_PRB12}. On the other hand, the AFM ordering in the iron telluride is bicollinear type with the ordering wave vector of ($\pi$/2,$\pi$/2) in contrast to the collinear AFM type with ($\pi$, 0) ordering in the iron arsenides. The band folding due to the ($\pi$, 0) ordering vector in the 122 materials results in the strong FS construction and the formation of new electron pockets~\cite{Terashima_PRL11} because the original electron and hole Fermi surfaces are folded onto each other, whereas strong band construction around $E_F$ in the 11 materials is not expected. The distinct orbital character in the electronic band structures near $E_F$ between 122 and 11 materials may be a possible origin of the opposite resistivity anisotropy. Note also that the bicolliner AFM ordering in the iron telluride results in the formation of zigzag Fe-Fe chains with the nearest neighbored spins aligning antiferromagnetically along the $b$ axis, whereas the AFM Fe-Fe chain in the iron pnictide is straight along the $a$ axis, as illustrated in Fig.~\ref{fig1}(a). The supposed impurity potential cloud appears to extend more in the AFM Fe-Fe direction in both systems. The recent STS study of Fe$_{1+x}$Te found the formation of an anisotropic electronic cloud with the triangular shape around the excess Fe atom  with the longer side along the $b$ axis~\cite{Machida_PRB13}. The interplay between the impurity and the specific electronic structure associated with the different spin configurations needs further investigation.


\section{summary}

A clear resistivity anisotropy was revealed for the doped FeTe systems when the temperature was decreased below $T_s$. Surprisingly, $\rho_a$ was always larger than $\rho_b$ in the parent compounds  Fe$_{1+x}$Te and Cu/Se-substituted crystals, which is opposite to the anisotropy observed in the iron arsenides. Since in the case of FeTe system the resistivity is larger in the direction of longer lattice spacing and AFM spin alignment, which is seemingly consistent with the naive picture, one may claim that the origin of the resistivity anisotropy is different between the two systems. However, it is found that the resistivity anisotropy in the iron telluride is mostly determined by the anisotropy in the RR component, and that the magnitude of resistivity anisotropy increased with increase of the amount of impurity atoms (while the AFM order is suppressed). It is suggested from our results that the resistivity anisotropy in the iron telluride is also induced by impurities, implying that around each excess Fe atom or substituted Cu/Se atom, an exotic anisotropic impurity state might be formed by anisotropically polarizing its electronic environment, and thereby acts as an anisotropic scattering center in the AFM state. The present results suggest that the impurity scenario may be the common phenomenological origin explicable for the resistivity anisotropy in both 11 and 122 compounds. Note that the polarization cloud formed around each impurity in  the iron telluride would be oriented along the crystallographic $b$ axis, whereas in the iron arsenides its orientation is along the $a$ axis, which deserves further observation using STS.


\section*{Acknowledgements} 

We thank D. Hirai and Y. Nakamura for their help in the measurements. L.L. thanks the Ministry of Education, Culture, Sports, Science, and Technology (MEXT) Scholarship of Japan and China Scholarship Council (CSC) for the financial support. This work was supported by the Transformative Research Project on Iron Pnictides (TRIP) from the Japan Science and Technology Agency, and by the Japan-China-Korea A3 Foresight Program from the Japan Society for the Promotion of Science (JSPS), and a Grant-in-Aid for Scientific Research from MEXT and JSPS.

\newpage


\newpage

\end{document}